# An Overview of Mobile Capacitive Touch Technologies Trends

Li Du, *Member IEEE*

baron1989113@gmail.com

*Abstract*—Touch sensing, as a major human/machine interface, is widely used in various commercial products such as smart watches, mobile phones, tablets and TVs. State-of-the-art touch detections are mainly based on mutual capacitive sensing, which requires necessary contact-touch, limiting the mobile user experience. Recently, remote gesture sensing is widely reported in both academy and industry as it can provide additional user-experience for mobile interface. The capacitive remote gesture sensing is mainly based on detecting self-capacitance, achieving high resolution through eliminating the parasitic mutual capacitance. In this work, we overview the different generations of touchscreen technology, comparing the touch and remote gesture sensing technologies difference. In addition, different remote gesture sensing technologies are also compared. The limitations and potentials of different topologies are discussed and a final conclusion about the technology trends is summarized in the end.

## I. INTRODUCTION

TOUCH SENSING, as a general HID, is widely implemented in various display products (e.g. smart watches, mobile phones, tablets, and TV). With the low power and light weight properties, modern touch sensing systems are gradually replacing the traditional interfaces devices. For example, it is rarely to see any mobile phone with a standby keyboard for typing or a tuning button for display brightness. The user-friendly interface provided by high resolutions touch sensor has opened doors for mobile OEM to develop touch related feature application such as mobile games and instant noting [1], further blooming the touch system market. According to a recent market report, there will be 2.8 billion touchscreen shipped to the market in 2016 [2]. The popularity of using touch screen in current commercial electronic devices has boosted the touch related research in both academy and industry. Among the reported works, majority of the work are about how to extract accurate finger position on the noisy tight touch space with relative low power consumption [3-6].

However, among those reported works, the touch sensing technology is still limited to two dimensional (2D) sensing, which requires the users to touch the screen directly for the system to determine finger position. This technology can inevitably bring several disadvantages; such as fingerprints on the screen or unresponsiveness with wet hands touch. The former case drives the users to put additional transparent plastic coverage on mobile screen [7] while the latter case force the mobile company to develop voice control which is not very mature in the current stage.

On the other side, advanced CMOS technology development has enabled to integrate more functions into one circuit system which cannot be realized several years ago. The so-called system-on-chip(SOC) solutions in the current circuit and system industry has brought many innovative applications such as RFI interconnect in the wireline communication [8-12], radar and navigation systems [13-15].

Drawbacks described above in the 2D sensing schemes and the fast growing CMOS technology have inspired the development of a SOC remote-sensing solution that will lead to three-dimensional (3D) gesture detection. Recently, a lot of publications report remote-sensing technology [16-21]. In general, they are separated into two categories. The most popular one is still using capacitive sensing, the other reported sensing methodology is using CMOS radar technology. However, it is still in the prototype stage [22]

In this paper, we overview the capacitive based touch sensing related technology. In Section II, conventional 2D capacitive sensing will be reviewed and Section III will be focused on introducing different technical approaches to achieve remote sensing through capacitive sensor. Section IV will compare different sensing methodology and conclusion will be drawn on Section V.

## II. CAPACITIVE TOUCH SENSING IN 2D

Projected capacitive touch (PCT) technology is regarded as the most popular capacitive sensing [23]. PCT touch screens are made up of a matrix of rows and columns of conductive electrodes. The two-layer electrodes are attached on the bottom side of the display glass. This can be done either by etching a single conductive layer to form a grid pattern of electrodes, or by etching two separate, perpendicular layers of conductive material with parallel lines or tracks to form a grid. The electrodes are usually fabricated as diamond shapes to track more finger-induced capacitance. Fig. 1 below is an example of this touch screen.

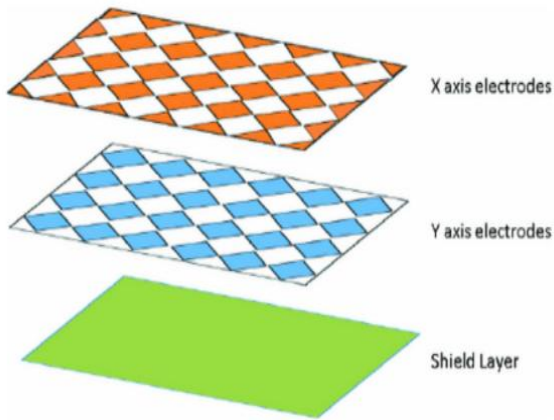

Fig. 1: Example of a PCT screen structure

The detection is through applying a voltage to this grid to create a uniform electrostatic field, which can be measured. When a conductive object, touching the PCT panel, it distorts the electrostatic field of the electrodes that are nearby the touch point. This is measurable as a change in the electrode's capacitance. If a finger bridges the gap between two of the electrodes, the charge field is further affected. The capacitance can be changed and measured at every individual point on the grid. Therefore, this system is able to accurately estimate the touch position. The changed capacitance can be detected through a high resolution capacitive sensing circuits. With proper processing, the touched finger position can be extracted through the capacitance information [24].

There are two types of PCTs: mutual-capacitive touch sensing and self-capacitive touch sensing. The widely used one is mutual capacitive sensing as it provides the capability of detecting multi-touch positions. Mutual-capacitive sensing is measuring the change of the coupling capacitance between two orthogonal electrodes and locating the finger position based on the changed electrodes' cross position shown as Fig. 2.

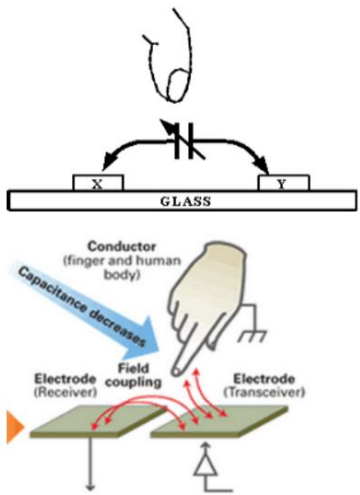

Fig. 2. Mutual-capacitive sensing methodology

The high resolution and multi-detection capability has made this type of touch sensing widely used in the industry. However, the large parasitic coupling capacitance is also limited this detection for remote sensing where the finger induced capacitance is much smaller than regular touched detection.

### III. CAPACITIVE BASED REMOTE SENSING

As described in Section II, the major drawback to implement remote sensing is the large coupling parasitic capacitance between electrodes to electrodes. To resolve this, Princeton researchers first propose Princeton researchers have proposed using self-capacitive sensing to detect small finger-induced capacitance, achieving a high resolution detection ability through using high-Q LC oscillator as the sensing blocks and customized large display panel. According to their report, the proposed architecture can achieve up to 30cm finger-height detection. Fig. 3 is shows a general diagram of the reported work [16].

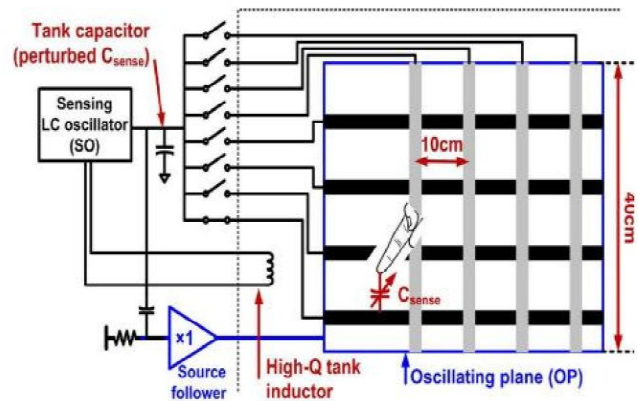

Fig. 3 3D Touch sensing system reported in ISSCC by Princeton's researchers.

Besides this work, UCLA researchers have also reported similar remote sensing system called Airtouch using self-capacitive sensing. Unlike the previous work, the Airtouch system is implemented directly on a mobile-sized touchscreen, making it potential to be integrated in the mobile devices. The channel coupling effect can be eliminated through creating a bootstrapped circuitry using a high gain operational amplifiers such as [25-26].

The relevant position estimation algorithm is also reported in their publication [19-20, 27]. The finger height detection range is up to 6cm with a power consumption of 2.3mW.

In addition to the academic report, industry company also gets attention on this new application. Specifically, Fogale Sensation [28-29] reported a remote sensing system on mobile phones through using two key technologies below:
1. Design bootstrap guards nearby the scanning channels to cancel interfering capacitance
2. A gird-level matrix electrodes array instead traditional two-dimension inter-cross sensor electrodes.

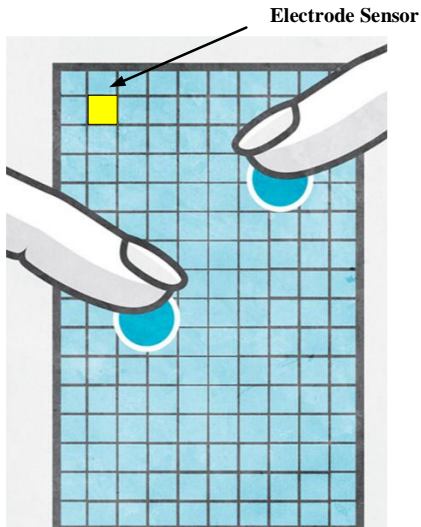

Fig. 4. Example of Fogale Technology Sensor Array

With the above two features, the Fogale's technology has been successfully demonstrated in the mobile phone. According to a report publish by Microsoft research team [30]. The near screen pre-touch motion can be detected and used in many innovative mobile applications.

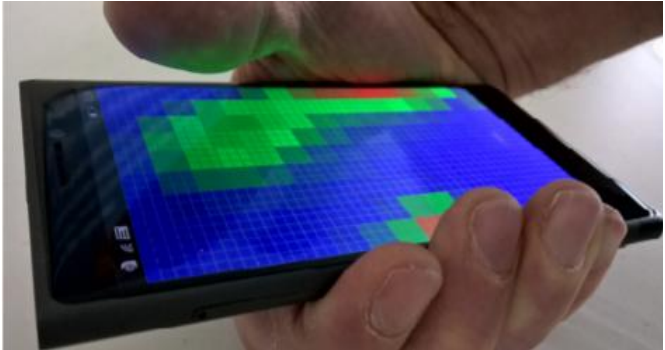

Fig. 5. Fogale Technology integrated with mobile phones, demonstrated by Microsoft Research Lab.

Another reported work in from Microchip. Inc, which first proposed a 3D touch sensing detection system as Ref [16]. The architecture of the proposed system is summarized as Fig. 2.10. The system used the TX to send out a low frequency AC signal and use the five RX to receive the capacitance change between the finger and the RX. By detecting the capacitance change between the finger and the five RX, the system reconstructs the finger position in space through comparing those response difference. The algorithm about finger reconstruction through the detected finger capacitance is not listed in any public materials.

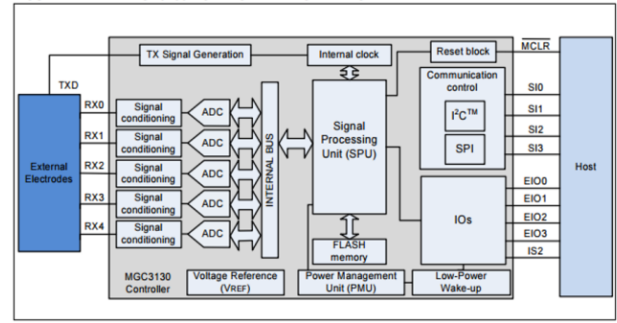

Fig. 6 System Diagram of the Microchip MGC3130.

To realize the high-sensitive capacitive sensing, the Microchip is also customizing the touch panel design. The touch panel pattern is shown in Fig. 7. According to its reported datasheet, the MGC3130 demonstrates a single-touch detection up to 15cm detection range with a spatial resolution up to 0.17mm and power consumption of 66mW in the acquisition mode.

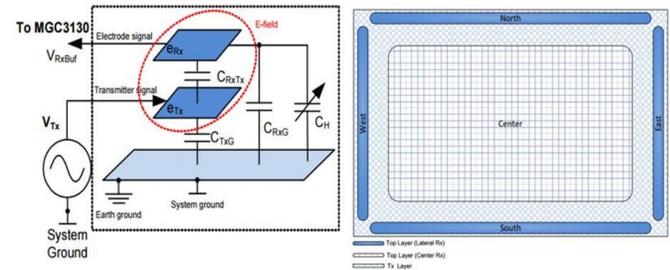

Fig. 7 Touch Panel Pattern of the Microchip MGC3130.

IV. COMPARISON OF REMOTE SENSING SOLUTIONS

Based on the literature review above, we summarize the following conclusions for the touch sensing technology. For the 2D regular touch sensor detection, the contact nature provides a large capacitive change in the touch channel, hence the system physical sensing resolution requirement is not high. There are multiple methods used in the industry and academy to achieve high resolution finger position detection. And the current most popular methodology is still using mutual capacitive sensing due to its compatibility with the smart phones' screens and the possibility of detecting multi-finger touch.

On the other side, 3D touch sensing is mainly based on the self-capacitive sensing to avoid mutual-capacitive affect. Among the reviewed work in Section III, all the demonstrated systems have used different analog circuits to cancel the inter-channel coupling capacitance. For instance, the Princeton researcher used an oscillating plane to isolate the coupling, the Fogale technology built a protection signal wall between electrodes to electrodes to eliminate this coupling effects and the UCLA researcher used a bootstrapped circuitry to achieve similar function.

In additional to eliminate the inter-channel coupling effects, one of the other shared features among the reported works is that the customized design of the touch panel. The widely used 2D touch panel is described in Section II. However, all the reported remote gesture sensing systems have redesigned the touch panel pattern. For instance, Fogale technology designed the touch panel as an electrodes matrix array with each electrode to be a square shape, while the UCLA and Princeton researchers reshape the electrodes to be either triangle or large-sized-electrode-distance.

Besides the sensing topology, the detection circuit also shows a similar trend. The Fogale technology didn't mention their detection circuits, however, both the UCLA and the Princeton researchers use similar sensing topology(Oscillator) to detect the varied self-capacitance in the sensor.

## V. CONCLUSION

In this paper, the capacitive based remote sensing technology is introduced. The current capacitive remote sensing is an evolution of the common touched detection, with an emphasis in solving the inter-channel coupling capacitance problem. This problem is solved through two approaches: One is through redesigning the touch panel pattern to increase the distance between electrodes to electrodes, while the other approach is to implement coupling cancellation circuit in the system to eliminate the inter-channel coupling.